----------
X-Sun-Data-Type: default
X-Sun-Data-Name: vortex.tex
X-Sun-Content-Lines: 401

\documentstyle[aps,preprint]{revtex}

\begin{document}

\draft

\title{\bf Electron localization by a magnetic vortex}

\author{R.~M.~Cavalcanti$^{a)}$}

\address{Departamento de F\'\i sica, Pontif\'\i cia Universidade
Cat\'olica do Rio de Janeiro \\ C.P.~38071, Rio de Janeiro, RJ, 
22452-970, Brasil}

\author{E.~S.~Fraga$^{b)}$ and C.~A.~A.~de Carvalho$^{c)}$}

\address{Instituto de F\'\i sica, Universidade Federal
do Rio de Janeiro \\ C.P.~68528, Rio de Janeiro, RJ, 21945-970, Brasil}

\date{\today}

\maketitle

\begin{abstract}

We study the problem of an electron in two dimensions in
the presence of a magnetic vortex with a step-like profile.
Dependending on the values of the effective mass and gyromagnetic
factor of the electron, it may be trapped by the vortex.
The bound state spectrum is obtained numerically, and
some limiting cases are treated analytically.

\end{abstract}

\newpage

Two-dimensional electronic systems with magnetic vortices
have become a topic of great interest, both
theoretical and experimental. In the theory of the fractional quantum 
Hall effect, it is common to resort to the  
composite fermion (electron $+$ magnetic flux) approach\cite{Hall};
a similar approach has been used in 
anyonic systems\cite{anyon}. In both cases, one is left, therefore, with a
problem of fermions moving in a random magnetic field generated
by flux tubes. On the experimental side, 
random magnetic fields have been realized by using 
a disordered type-II superconductor in a uniform magnetic field
as the substrate for a 2D electron gas: the induced flux lines are trapped 
by the defects in the 
superconductor, generating a random array of vortices\cite{SUC}. 
A demagnetized ferromagnet with randomly oriented magnetic
domains has also been used as a substrate\cite{FM}.
Therefore, the study of the influence of one or more vortices on 
the spectrum of a two-dimensional electron gas is a very important 
issue, especially with respect to the eventual presence of localized states.

In this work, we study the problem of an electron in two
dimensions in the presence of a magnetic flux (a vortex).
The system is described by a Pauli Hamiltonian, and the vortex
has the simplest profile, i.e., $B(r)=B_V \theta(a-r)$, where
$\theta(x)$ is the Heaviside step function and $a$ defines the radius
of the vortex. The effective gyromagnetic factor may differ from $g=2$. 
In fact, although the Dirac equation predicts this value, {\it QED} radiative
corrections\cite{Schwinger} give for the electron $g$-factor a value  
slightly bigger than $2$ ($g_{QED}=2.0023$), a result confirmed
(in fact, discovered) by experiment\cite{Foley}. In solids,
the $g$-factor of the electron may depart significantly from 2
as a result of spin-orbit coupling effects\cite{Callaway} (for
instance, $g=0.32$ in $n$-GaAs and $-44$ in $n$-InSb). However,
as we shall see shortly, what determines the existence
of bound states is the quantity $gm^*/m\equiv g^*$, where $m^*$
is the effective mass of the electron ($g^*=0.026$ in $n$-GaAs and $-0.66$ 
in $n$-InSb \cite{Seeger})

The Hamiltonian has the following form:
\begin{equation}
\label{Pauli}
H=\frac{1}{2m^*}\left( -i\hbar \nabla-\frac{e}{c}{\bf A}
\right)^2 - \frac{1}{2}\,g\mu_B\sigma_zB,
\end{equation}
where $\mu_B=e\hbar/2mc$ is the Bohr magneton, and
$\sigma_z=diag(1,-1)$ is the third Pauli matrix. The magnetic field
points in the $z$-direction.
In units such that $2m^*=\hbar=c=1$, this 
simplifies to
\begin{equation}
\label{Pauli2}
H=\left( -i\nabla-e{\bf A}\right)^2 - \frac{1}{2}g^*eB\sigma_z.
\end{equation}

In the case of a magnetic field with radial symmetry, one can
choose the vector potential
in the symmetric gauge, ${\bf A}=A(r)\,{\bf e}_{\phi}$, so that
the magnetic field is written as
$B=\frac{1}{r}\,\frac{d}{dr}\,[rA(r)]$, and
Eq.~(\ref{Pauli2}) becomes
\begin{equation}
\label{Pauli3}
H=-\frac{1}{r}\,\frac{\partial}{\partial r}\left(r\,\frac{\partial}
{\partial r}\right) - \frac{1}{r^2}\,\frac{\partial^2}
{\partial\phi^2} + 2ieA\,\frac{1}{r}\,\frac{\partial}{\partial\phi} +
e^2A^2-\frac{1}{2}\,g^*eB\sigma_z
%\equiv\pmatrix{H_{-} & 0\cr 0 & H_{+}\cr}.
\end{equation}
Thus, writing the eigenfunctions of $H$ as
$\psi(r,\phi)=u_{\ell}(r)\,e^{i\ell\phi}$, 
we obtain the eigenvalue equation
\begin{equation}
\label{eigeneq}
\left( -\frac{d^2}{dr^2}-\frac{1}{r}\,\frac{d}{dr}
+ \frac{\ell^2}{r^2}-2\ell eA\,\frac{1}{r} + e^2A^2
- \frac{1}{2}\,g^*eB\sigma_z\right)  u_{\ell}(r)=E\, u_{\ell}(r),
\end{equation}
which is equivalent to the pair of equations
\begin{equation}
\label{H+-}
H_{\pm}u_{\ell}^{(\pm)}(r)\equiv
\left( -\frac{d^2}{dr^2}-\frac{1}{r}\,\frac{d}{dr}
+ \frac{\ell^2}{r^2}-2\ell eA\,\frac{1}{r} + e^2A^2
\pm \frac{1}{2}\,g^*eB\right)  u_{\ell}^{(\pm)}(r)
=E_{\pm}\, u_{\ell}^{(\pm)}(r).
\end{equation}
The $-(+)$ sign corresponds to spin-up(down) electrons.

Now, let us consider the field of a vortex, $B=B_V\theta(a-r)$.
The vector potential corresponding to this field is
\begin{equation}
A(r)=\cases{\frac{1}{2}\,B_Vr & for $r<a$, \cr
            B_Va^2/2r & for $r>a$. \cr}
\end{equation}
Since we are interested in the possible existence of
bound states, we shall consider only the spin-up sector of
Eq.~(\ref{H+-}) (in the spin-down sector, the coupling
of the spin with the field gives rise to a {\em repulsive}
potential). For $E_{-}\equiv-\mu^2<0$, this equation has 
the following solution (inside and outside the vortex of radius $a$)
\begin{mathletters}
\begin{eqnarray}
u_{in}(r)&=&C_{in}\,e^{-\Phi\,r^2/2a^2}\,(r/a)^{|\ell|}\,
M\left(\alpha,|\ell|+1,\Phi\,r^2/a^2 \right), \\
u_{out}(r)&=&C_{out}\,K_{|\ell-\Phi|}(\xi r/a), \label{u_out}
\end{eqnarray}
where $\Phi \equiv eB_V a^2/2$ is the magnetic flux in units
of the quantum flux (here we assume $\Phi$ positive), $\xi \equiv \mu a$,
\end{mathletters}
\begin{equation}
\alpha \equiv \frac{\xi^2}{4\Phi}-\frac{1}{2}\left(\ell-|\ell|-1+
\frac{g^*}{2}\right),
\end{equation}
$M(a,b,z)$ is the Kummer (or confluent hypergeometric) function
and $K_{\nu}(z)$ is the modified Bessel function\cite{Abramowitz}.

In order to find the energies of the bound states, one must
impose continuity of the logarithmic derivative of $u(r)$ at
the boundary of the vortex, or, equivalently, 
look for the zeroes of the function
\begin{eqnarray}
R(\xi,\Phi,\ell,g^*)&\equiv& a\left[\frac{u'_{out}(r)}{u_{out}(r)}-
\frac{u'_{in}(r)}{u_{in}(r)}\right]_{r=a} \nonumber \\
&=&|\ell-\Phi|-\frac{\xi K_{|\ell-\Phi|+1}(\xi)}
{K_{|\ell-\Phi|}(\xi)}+\Phi-|\ell|-\frac{2\alpha\Phi}{|\ell|+1}\,
\frac{M(\alpha+1,|\ell|+2,\Phi)}{M(\alpha,|\ell|+1,\Phi)}.
\label{zeroes}
\end{eqnarray}
Now, we shall analyze (\ref{zeroes}) numerically for representative values of 
the parameters, and consider the analytical solution of some limiting 
cases to provide some physical insight.

For $g^*=2$, a theorem due to Aharonov and Casher 
\cite{Aharonov} tells us that there exists $N$ zero-energy
normalizable states, where $N$ is an integer defined through
$\Phi=N+\epsilon$, $0<\epsilon\le 1$. Let us
show this explicitly: taking $\xi\to 0$ in
Eq.~(\ref{zeroes}), the last term vanishes if $\ell\ge 0$,
and the second term tends to $-2\,|\ell-\Phi|$. Therefore,
\begin{equation}
R(0,\Phi,\ell,2)=-|\ell-\Phi|+\Phi-\ell,
\end{equation}
which vanishes for $\ell\le\Phi$. 
There is a further restriction on the values of $\ell$. 
When $\xi=0$, the solution
outside the vortex, Eq.~(\ref{u_out}), must be replaced by
\begin{equation}
u_{out}(r;\xi=0)=C_{out}\,r^{-|\ell-\Phi|}.
\end{equation}
In order for $u_{\ell}(r)$ to be square-integrable, we need
$|\ell-\Phi|>1$. Together with the condition
$0\le\ell\le\Phi$ and the definition of $N$,
we obtain the following inequality:
$0\le\ell<\Phi-1\le N$.
Thus, except for a normalization factor,
the corresponding radial wave functions are
\begin{equation}
u_{\ell}^{(-)}(r)=\cases{e^{-\Phi\,r^2/2a^2}\,(r/a)^{\ell} & for $r<a$, \cr
                         e^{-\Phi/2}\,(r/a)^{\ell-\Phi}  & for $r>a$. \cr}
\qquad(\ell=0,1,\ldots,N-1)
\end{equation}
A remark is in order here: such states can only exist if $a\ne 0$.
For $a\to 0$, they develop a non-integrable singularity at the
origin.

If $g^*>2$, we can find eigenstates with negative energy.
In order to create a bound state with angular momentum $\ell$,
a minimum value of $\Phi$ is required;
let us call it $\Phi_{\ell}$ (clearly, for $\Phi>\Phi_{\ell}$, this
bound state will be pushed to negative energy). 
To find its dependence on $\ell$,
one must solve the equation $R(0,\Phi_{\ell},\ell,g)=0$, which,
for $\ell\ge 0$, can be written as
\begin{equation}
\label{Phi_l}
-|\ell-\Phi_{\ell}|+\Phi_{\ell}-\ell-
\frac{2\alpha_0\Phi_{\ell}}{\ell+1}\,
\frac{M(\alpha_0+1,\ell+2,\Phi_{\ell})}
{M(\alpha_0,\ell+1,\Phi_{\ell})}=0,
\end{equation}
with $\alpha_0\equiv\alpha(\xi=0)=(2-g^*)/4$. For $\ell=0$,
it has $\Phi_0=0$ as solution. For $\ell>0$, Eq.~(\ref{Phi_l})
is too complicated to be solved analytically, but it 
becomes tractable if $g^*-2\ll 1$. In fact, assuming 
that $\Phi_{\ell}=\ell-\delta$, with $\delta\ll 1$,
and expanding Eq.~(\ref{Phi_l})
to first order in small quantities, one finds
\begin{equation}
\label{Phiell}
\Phi_{\ell}\approx \ell\left[1-\frac{g^*-2}{4(\ell+1)}\,M(1,\ell+2,\ell)
\right].
\end{equation}
Note that there is a discontinuity at $g^*=2$, which is not apparent
in the expression above. 
In fact $\Phi_{\ell}(g^*=2)=\ell+1$, and {\em not} $\ell$, 
as sugested by the $g^* \to 2$ limit of (\ref{Phiell}).
This happens because even for an infinitesimal binding energy the
wavefunction has an exponential decay outside the vortex,
thus eliminating the requirement that $|\ell-\Phi|$ be greater
than one to guarantee the normalizability of the wavefunction.
The behavior of $\Phi_{\ell}$ as a function of $g^*$, for different
values of angular momentum is shown in Fig.~1. 

Now, let us examine how the energy of the bound state with
$\ell=0$ depends on the magnetic flux. Assuming that
$\xi\ll 1$ for $\Phi\ll 1$, we can approximate $R(\xi,\Phi,\ell=0,g^*)$ by
\begin{equation}
2\Phi\left\{1-\left[1-\frac{\Gamma(1-\Phi)}{\Gamma(1+\Phi)}\,
\left(\frac{\xi}{2}\right)^{2\Phi}\right]^{-1}-\alpha_0\left[1-\frac{1}{2}\,
(1-\alpha_0)\,\Phi\right]\right\}.
\end{equation}
In this approximation, solving $R(\xi,\Phi,0,g^*)=0$ for $\xi$ we find
\begin{equation}
\label{xi}
\xi(\Phi,\ell=0,g^*)=2\left[\frac{\Gamma(1+\Phi)}{\Gamma(1-\Phi)}
\left(\frac{g^*-2}{g^*+2}\right)\left(\frac{1+\frac{1}{8}\,(g^*+2)\,\Phi}
{1+\frac{1}{8}\,(g^*-2)\,\Phi}\right)\right]^{1/2\Phi}.
\end{equation}
Since we are assuming that $\Phi\ll 1$, we can write 
$\Gamma(1\pm\Phi)\approx 1\mp\gamma\Phi$ (where $\gamma=0.577\ldots$
is Euler's constant) \cite{Abramowitz} and use
$\lim_{x\to 0}(1+x)^{1/x}=e$ to simplify (\ref{xi}) to
\begin{equation}
\label{approx}
\xi(\Phi,0,g^*)\approx 2\,e^{-\gamma+1/4}\left(\frac{g^*-2}{g^*+2}
\right)^{1/2\Phi}.
\end{equation}
The full spectrum may be obtained numerically for particular values of
$g^*$, for different values of the angular momentum. Illustrative
results are presented in Fig.~2. For the sake of comparison, 
the approximate result for
$\xi(\Phi,0,g^*)$, Eq.~(\ref{approx}), is plotted with the
exact result in Fig.~3.  

For $g^*>6$, Eq.~(\ref{Phi_l}) may posses more
than one solution, each one corresponding to a different bound state.
This happens because $\alpha_0+1$ becomes negative, so that
$M(\alpha_0+1,\ell+2,\Phi_{\ell})$ has at least one non-zero 
root\cite{Kummer}; if such a root is larger than $\ell$, it
is also a solution of Eq.~(\ref{Phi_l}). Since, as far as we know,
such large values of $g^*$ are not found in Nature,
we shall not analyze them further.

The authors acknowledge CNPq, CAPES, FINEP and FUJB/UFRJ for financial 
support.

%%%%%%%%%%%%%%%%%%%%%%%%%%%%%%%%%%%%%%%%%%%%%%%%%%%%%%%%%%%%%%%%%%%%%%%%

%\end{document}

\newpage

%%%%%%%%%%%%%%%%%%%%%%%%%%%%%%%%%%%%%%%%%%%%%%%%%%%%%%%%%%%%%%%%%%%%%%%%

% FIGURES
% -------

\noindent
\underline{\bf Figure Captions}:

\vspace{5mm}
\noindent
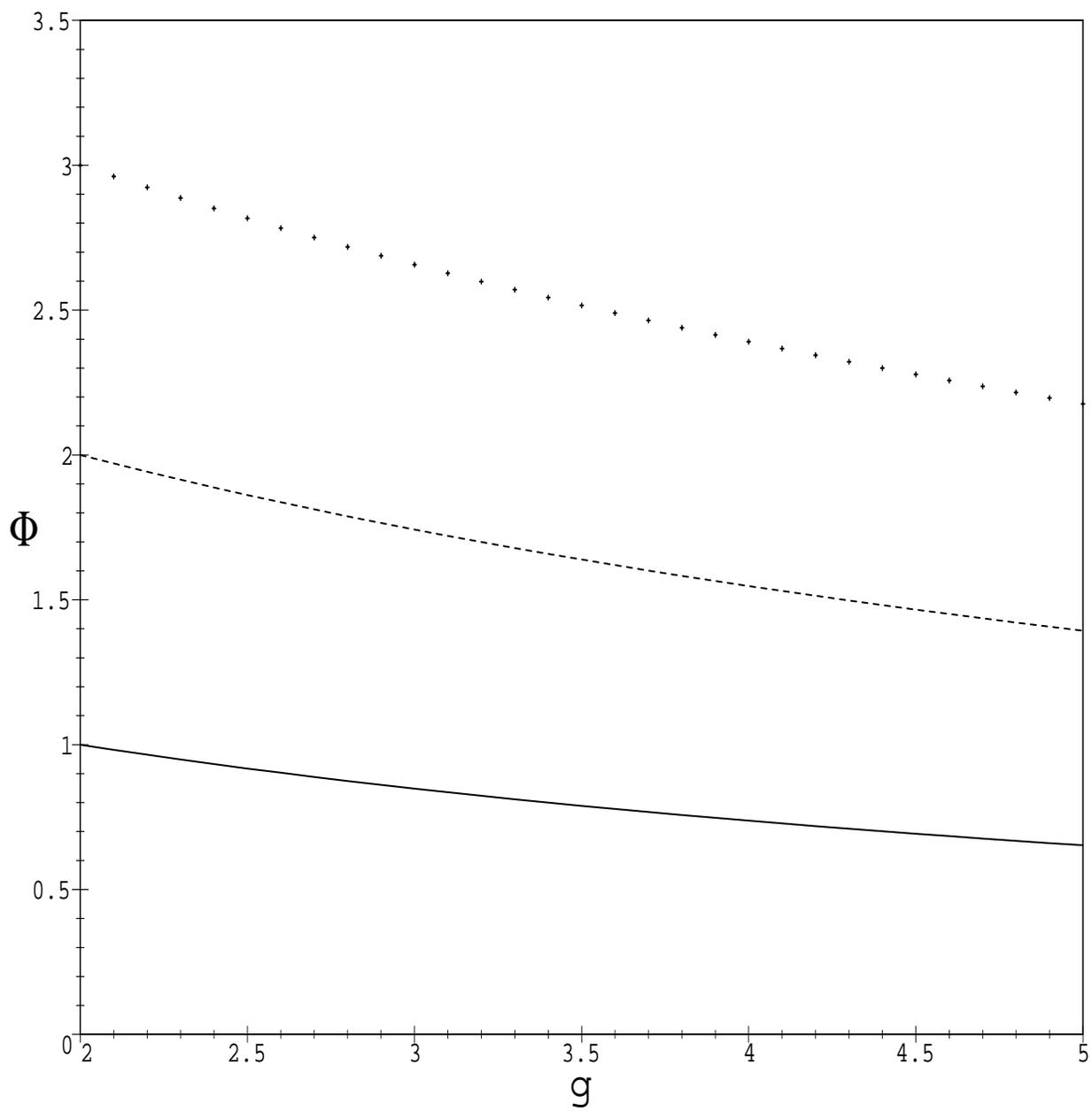
{\bf Figure 1}: $\Phi_{\ell}$ vs.\ $g^*$ for $\ell=1$ (solid line),
$\ell=2$ (dashed line) and $\ell=3$ (crosses). (For $g^*=2$, see 
remark in the text.)

\vspace{5mm}
\noindent
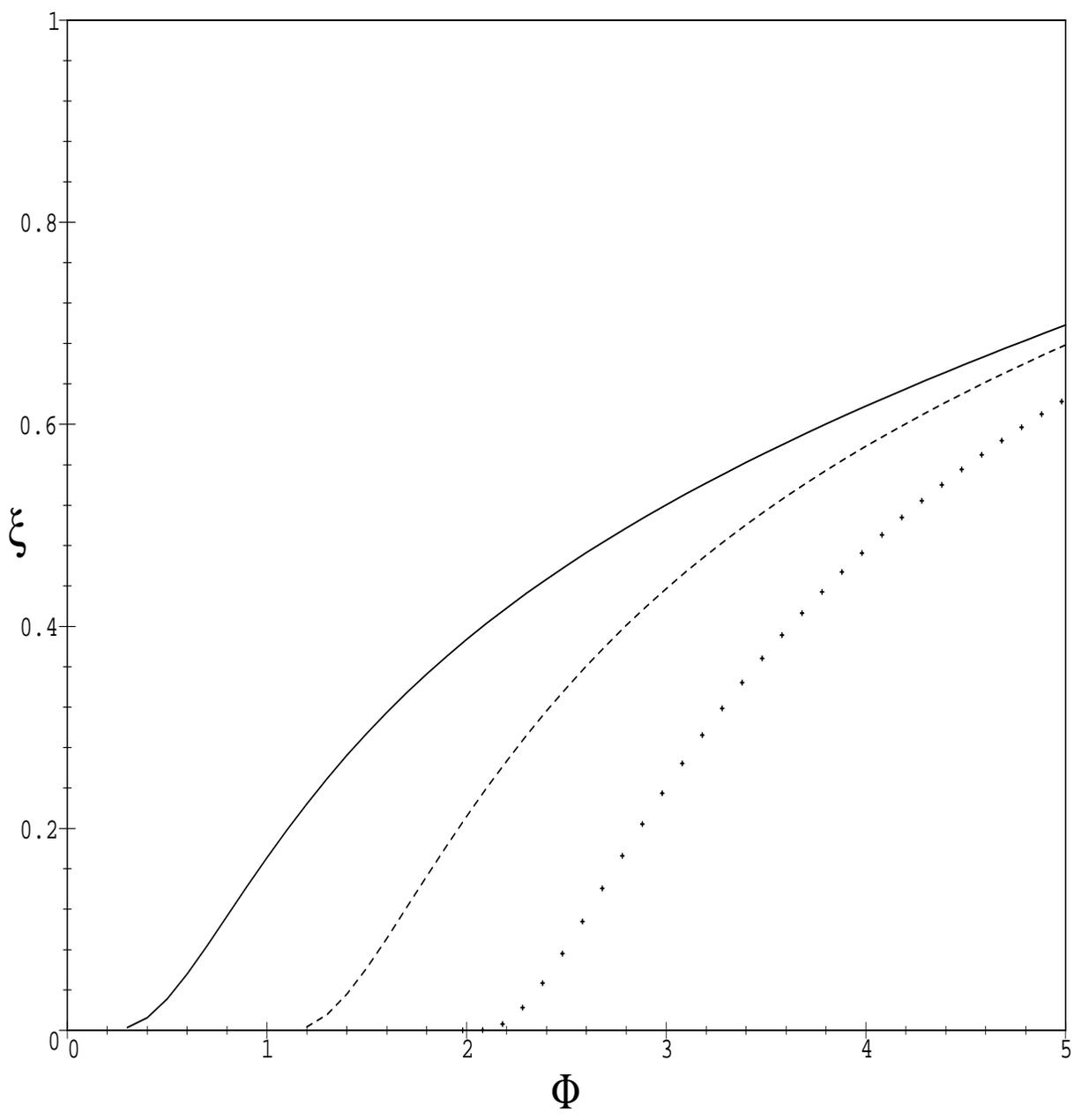
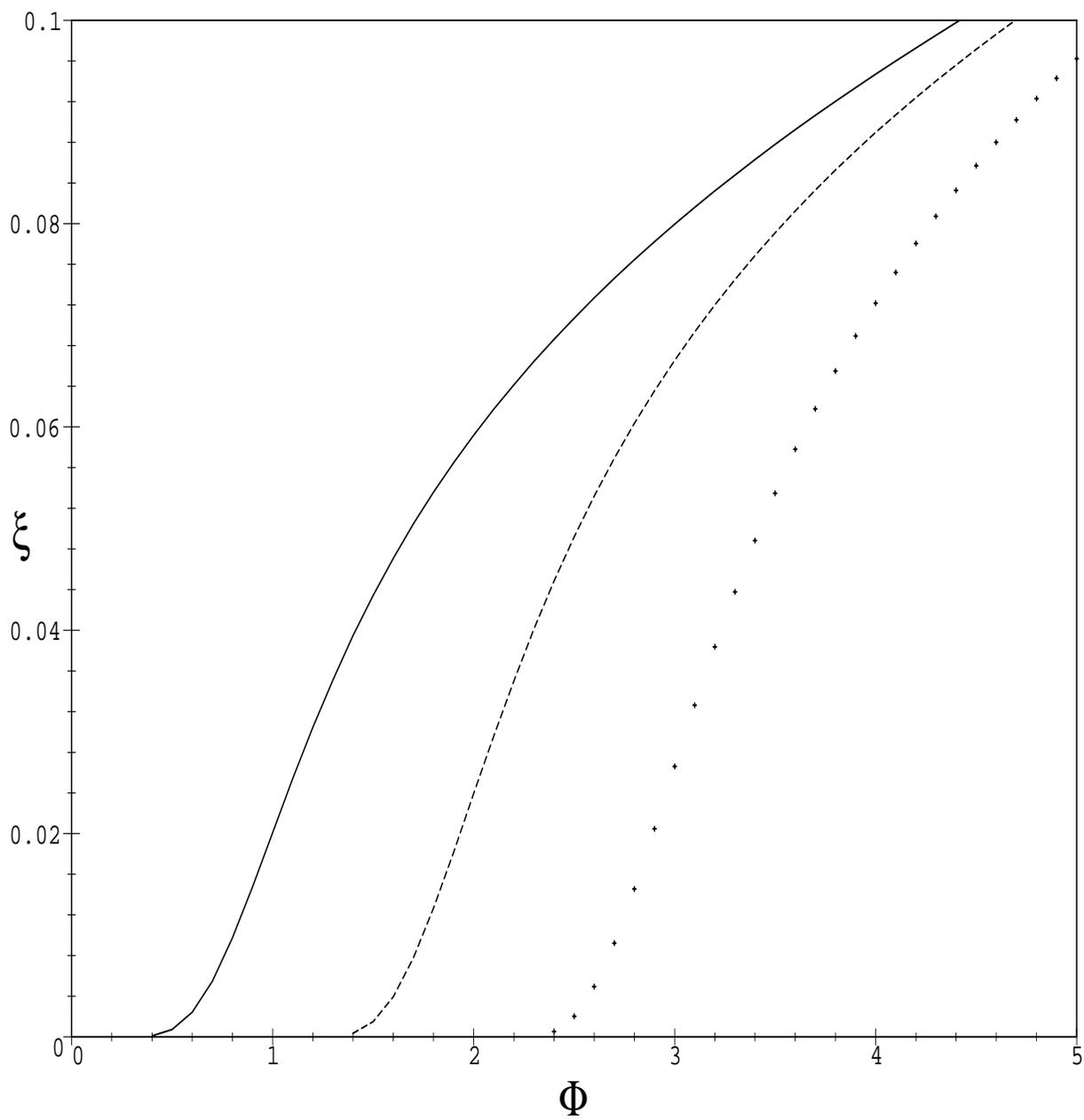
{\bf Figure 2}: $\xi$ vs.\ $\Phi$ for $\ell=0$ (solid line), 
$\ell=1$ (dashed line), and $\ell=2$ (crosses): (a) $g=2.1$,
(b) $g=2.0023$.

\vspace{5mm}
\noindent
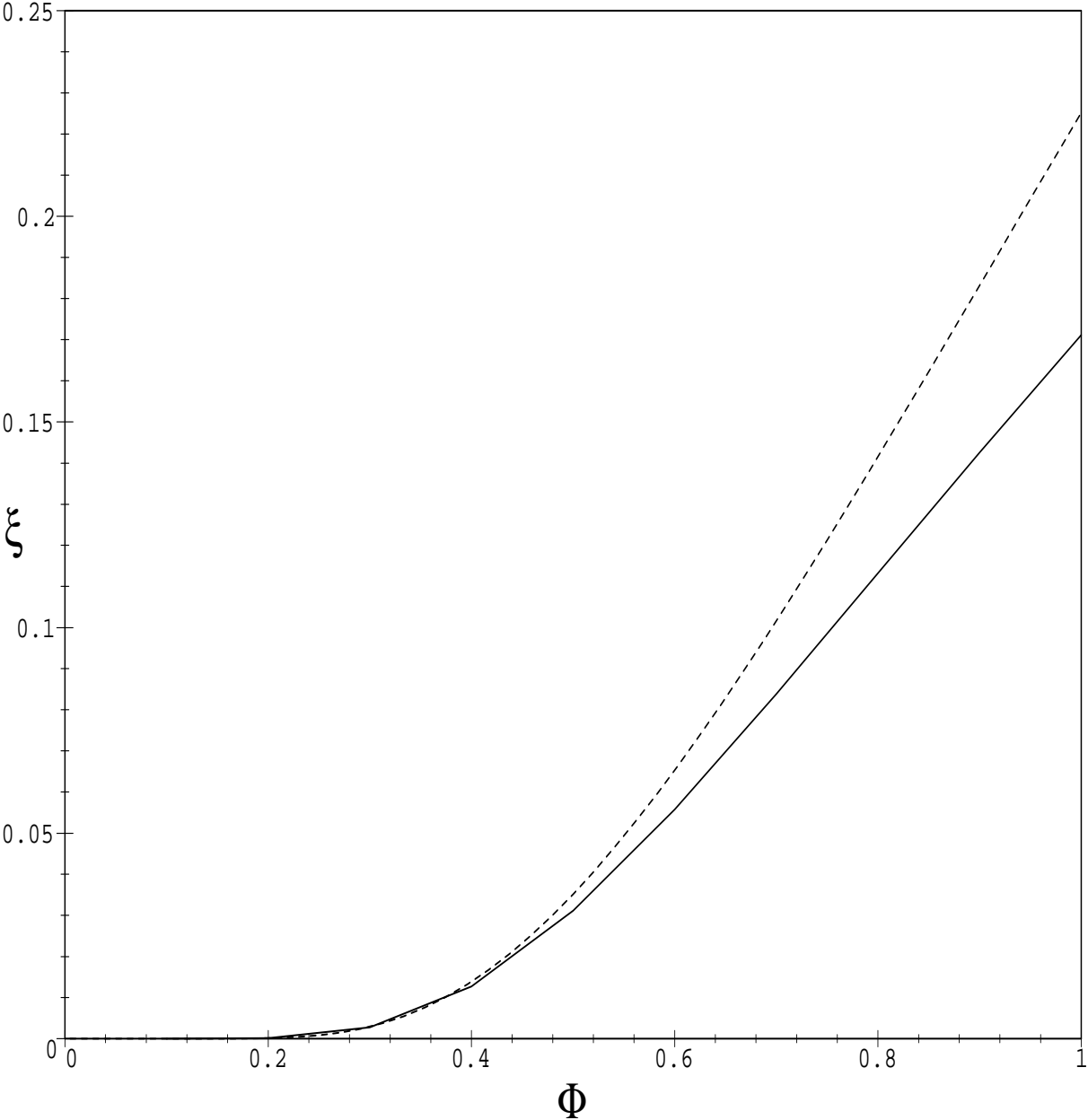
{\bf Figure 3}: $\xi$ vs.\ $\Phi$ for $g=2.1$ and $\ell=0$:
exact (solid line) and approximate, as given by Eq.~(\ref{approx}) (dashed
line).

\end{document}